\documentstyle[axodraw]{article}

\def\ltap{~\raisebox{-.4ex}{\rlap{$\sim$}} \raisebox{.4ex}{$<$}~}
\def\gtap{~\raisebox{-.4ex}{\rlap{$\sim$}} \raisebox{.4ex}{$>$}~}
\newcommand{\Rsl}{{\not\!{R}}}
\renewcommand{\thefootnote}{\fnsymbol{footnote}}

\title{\begin{flushright}
\small May 1998 \hfill SINP/TNP/98-13\\
{\tt hep-ph/9806214} 
\end{flushright}
{\Large\bf New constraints on $R$-parity violation from proton 
stability}}

\author{
{\bf Gautam Bhattacharyya}%
\thanks{E-mail address: gb@tnp.saha.ernet.in}~~and 
{\bf Palash B. Pal}%
\thanks{E-mail address: pbpal@tnp.saha.ernet.in} \\ 
{\em Saha Institute of Nuclear Physics, 1/AF Bidhan Nagar, 
Calcutta 700064, India}
}

\date{}

\begin{document}

\maketitle

\begin{abstract}

{\small We derive stringent upper bounds on all the $(\lambda''_{ijk}
\mu_l)$-type combinations from the consideration of proton stability,
where $\lambda''_{ijk}$ are baryon-number-violating trilinear
couplings and $\mu_l$ are lepton-number-violating bilinear mass
parameters in a $R$-parity-violating supersymmetric theory.}

\end{abstract}

\vskip 20pt  

\setcounter{footnote}{0}
\renewcommand{\thefootnote}{\arabic{footnote}}

In the standard model (SM) of particle interactions, baryon and lepton
numbers ($B$ and $L$ respectively) appear as accidental global
symmetries which are not violated in any order of perturbation
theory. On the other hand, in the minimally supersymmetrized version
of this model (MSSM), unless one {\em assumes}\/ that those are indeed
conserved quantities, $R$-parity-violating ($\Rsl$) couplings are
naturally allowed \cite{rpar}. Defined as $R = (-1)^{(3B+L+2S)}$
(where $S$ is the spin of the particle), $R$-parity is a discrete
symmetry under which SM particles are even while their superpartners
are odd. In the MSSM, the most general superpotential is expressed as
	\begin{equation} 
W = W_0 + W',
	\end{equation} 
where $W_0$ and $W'$ represent the $R$-parity-conserving and
$R$-parity-violating interactions, respectively. In terms of the MSSM
superfields, $W_0$ and $W'$ can be expressed as  
	\begin{eqnarray} 
\label{W0}
W_0 & = & f_e^{ij} L_i H_d E_j^c + f_d^{ij} Q_i H_d D_j^c
+ f_u^{ij} Q_i H_u U_j^c + \mu H_d H_u, \\
\label{W'}		
W' & = & {1\over{2}}\lambda_{ijk} L_i L_j E^c_k + 
\lambda'_{ijk} L_i Q_j D^c_k + 
{1\over{2}}\lambda''_{ijk} U^c_i D^c_j D^c_k + 
\mu_i  L_i H_u,
	\end{eqnarray}
where $i$, $j$, $k$ are generation indices, which are assumed to be
summed over. In eqs.~(\ref{W0}) and
(\ref{W'}), $L_i$ and $Q_i$ are
SU(2)-doublet lepton and quark superfields, $E^c_i,
U^c_i, D^c_i$ are SU(2)-singlet charged lepton, up- and down-quark
superfields, and  $H_d$ and $H_u$ are Higgs superfields that
are responsible for the down- and up-type masses 
respectively. In eq.~(\ref{W'}), $\lambda$, $\lambda'$ and
$\mu_i$ are $L$-violating while $\lambda''$ are $B$-violating
parameters. $\lambda_{ijk}$ is antisymmetric
under the interchange of the first two family indices, while
$\lambda''_{ijk}$ is antisymmetric under the interchange of the last
two. Thus there could be 27 $\lambda'$-type, 9 each of $\lambda$-
and $\lambda''$-type couplings and 3 $\mu_i$ parameters when
$R$-parity is explicitly broken.

Since no evidence of $L$ or $B$ violation has been found in
experiments to date, one can take either of the two attitudes. The
first is to impose $R$-parity that forbids all the terms in $W'$. The
second is to take the view that since there is no good theoretical
motivation for applying such a symmetry {\em a priori}, perhaps it is
more general to admit all interactions invariant under supersymmetry
and gauge symmetry, and to examine what are the bounds on the
couplings in $W'$ from various phenomenological considerations
\cite{review}. We take the latter approach in this paper.

The prime phenomenological concern attached to any $\Rsl$ theory is
the question of proton stability. Proton decay requires {\em both}\/ $L$
and $B$ violations. Therefore the nonobservation of proton decay could
be translated to bounds on the simultaneous presence of $L$- and
$B$-violating terms. Thus what could be obtained in this way are
correlated bounds involving the $\lambda''$ with one of the
$\lambda'$, $\lambda$ or $\mu_i$. Investigations along this direction
have already revealed the following constraints. 

($i$) The simultaneous presence of $\lambda'$ and $\lambda''$
couplings involving the lighter generations drives proton decay ($p
\rightarrow \pi^0 e^+$) at tree level yielding extremely tight
constraints. From the bound on the proton lifetime $\tau_p \gtap
10^{32}$\,y in the given channel, one obtains \cite{HK}, for an
exchanged squark mass of 1 TeV,
	\begin{equation} 
\lambda'_{11k} \lambda''_{11k} \ltap 10^{-24},
\label{lpldp}
		\end{equation}
where $k = 2, 3$. These constraints weaken linearly with the squark
mass as the latter increases.

($ii$) One can always find at least one diagram at one loop level in
which any $\lambda'_{ijk}$ in conjunction with any $\lambda''_{lmn}$
contributes to proton decay \cite{SmVi98}. It follows that for an
exchanged scalar mass of 1 TeV, 
	\begin{equation}
\lambda'_{ijk} \lambda''_{lmn} \ltap 10^{-9}.  
	\end{equation}
If one admits tree level flavour-changing squark mixing, the bounds
are strengthened by two orders of magnitude. 

($iii$) The combination of $\lambda''$ and $\lambda$, in association
with the charged lepton Yukawa coupling $f_e$, drives
$p \rightarrow K^+ \ell^{\pm} \ell'^{\mp}\bar{\nu}$ at tree level
\cite{LoPa97}. It should be noted though that
the analysis in 
\cite{LoPa97} is based on an extended ${\rm SU(3)}_{\rm c} \times {\rm
SU(3)}_{\rm L} \times {\rm U(1)}$ gauge model. However the argument
applies also to the usual ${\rm SU(3)}_{\rm c}
\times {\rm SU(2)}_{\rm L} \times {\rm U(1)}$ gauge theory, and the
same bound holds. In absence of any direct experimental limits on the
above decay modes, this bound is 
	\begin{equation} 
\lambda''_{112}\lambda_{ijk} \ltap 10^{-16} \qquad (k\neq 3) \,,
	\end{equation} 
assuming a benchmark
value $\tau_p \gtap 10^{31}~y$ and an exchanged scalar mass of 1 TeV.

\begin{figure}
\begin{center}
\begin{picture}(190,100)(0,0)
\SetWidth{1.2}
\ArrowLine(10,10)(50,50)
\Text(25,30)[r]{$d^c$}
\ArrowLine(10,90)(50,50)
\Text(25,70)[r]{$s^c$}
\DashArrowLine(100,50)(50,50){3}
\Text(75,55)[b]{$\widetilde u^c$}
\Text(51,40)[bl]{$\lambda''_{112}$}
\ArrowLine(100,50)(100,90)
\Text(97,70)[r]{$q_1$}
\ArrowLine(100,50)(140,50)
\Text(120,55)[b]{$H_u$}
\ArrowLine(180,50)(140,50)
\Text(165,55)[b]{$\psi_l$}
\Text(100,40)[b]{$f_u$}
\Text(140,40)[b]{$\mu_l$}
\SetWidth{1.2}
\Line(135,45)(145,55)
\Line(135,55)(145,45)
\end{picture}
\end{center}
\caption[]{\small\sf Proton decay mediated by the $B$-violating
$\lambda''_{112}$ in association with the $L$-violating
$\mu_l$ at the tree level. $\psi_l$ is the lepton
doublet.}\label{f:pdk} 
\end{figure}
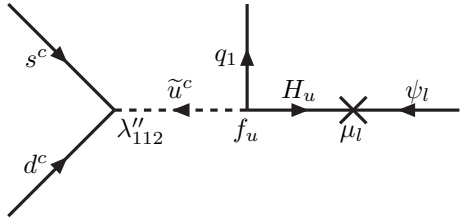
What therefore remains to be done is to examine the bounds on the
third and final type of $L$- and $B$-violating product couplings
$\lambda''_{ijk}\mu_l$. This is precisely our goal in the present
paper.

Consider the mechanism of nucleon decay shown in
Fig.~\ref{f:pdk}. Here one 
utilizes the $B$-violating $\lambda''_{112}$, the $L$-violating
$\mu_l$ and the canonical $R$-parity-conserving Yukawa coupling 
$f_u$. For the sake of convenience, we assume that all flavour indices
correspond to particles in their mass basis. Recalling that
$\lambda''_{ijk}$ is antisymmetric under the interchange of the last
two indices, at the quark level we obtain the processes
	\begin{equation}
d^c s^c \rightarrow u \bar{\nu}_l, \qquad 
d^c s^c \rightarrow d \bar{l} \,. 
	\end{equation}
They imply the following decay modes of the
proton: 
	\begin{equation} 
p \rightarrow K^+ \nu_l, ~~~~ p \rightarrow K^+ \pi^+ l^- \,,
\label{dmode} 
	\end{equation} 
where the charged lepton in the final state can only be $e$ or 
$\mu$. 
The relevant dimension-6 operators have an effective coupling
	\begin{equation} 
G_{\Rsl} \simeq \frac{\lambda''_{112} f_u}{m^2_{\tilde{u}_R}} 
                \left(\frac{\mu_l}{m_{\tilde{H}_u}}\right), 
	\end{equation}
leading to an approximate proton lifetime
	\begin{equation} 
\tau_p^{\Rsl} \simeq \left(m_p^5 G^2_{\Rsl}\right)^{-1}.
	\end{equation}
Among the decay modes in eq.~(\ref{dmode}), the channel $p\to K^+\nu$
($\tau_p\gtap 10^{32}\,y$ \cite{pdg}) offers the most stringent
constraints.  For a squark mass $m_{\tilde{u}_R} = 1$\,TeV, we obtain 
	\begin{equation}
\lambda''_{112} \epsilon_l \ltap 10^{-21},
	\end{equation}
where $\epsilon_l$ is defined through $\mu_l \equiv \epsilon_l
\mu$. For our order of magnitude estimate, we have made a
simple approximation $\mu \simeq m_{\tilde{H}_u} \simeq 1~{\rm TeV}$.

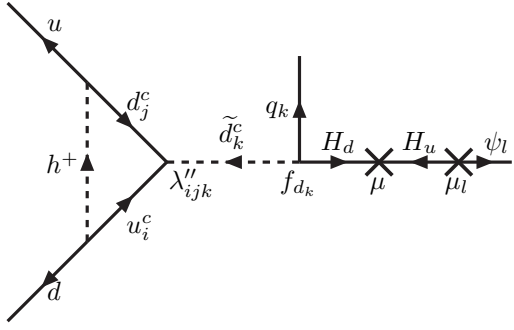
\begin{figure}
\begin{center}
\begin{picture}(200,100)(-10,0)
\SetWidth{1.2}
\ArrowLine(20,20)(50,50)
\Text(35,30)[tl]{$u^c_i$}
\ArrowLine(20,20)(-10,-10)
\Text(5,5)[tl]{$d$}
\ArrowLine(20,80)(50,50)
\Text(35,70)[bl]{$d^c_j$}
\ArrowLine(20,80)(-10,110)
\Text(5,100)[bl]{$u$}
\DashArrowLine(20,20)(20,80){3}
\Text(17,50)[r]{$h^+$}
\DashArrowLine(100,50)(50,50){3}
\Text(75,58)[b]{$\widetilde d^c_k$}
\Text(51,40)[bl]{$\lambda''_{ijk}$}
\ArrowLine(100,50)(100,90)
\Text(97,70)[r]{$q_k$}
\Text(100,40)[b]{$f_{d_k}$}
\ArrowLine(100,50)(130,50)
\Text(115,55)[b]{$H_d$}
\ArrowLine(160,50)(130,50)
\Text(146,55)[b]{$H_u$}
\ArrowLine(160,50)(180,50)
\Text(175,55)[b]{$\psi_l$}
\Text(130,40)[b]{$\mu$}
\Text(160,40)[b]{$\mu_l$}
\SetWidth{1.4}
\Line(125,45)(135,55)
\Line(125,55)(135,45)
\Line(155,45)(165,55)
\Line(155,55)(165,45)
\end{picture}
\end{center}
\caption[]{\small\sf Loop-driven proton decay mediated by the
$B$-violating 
$\lambda''_{ijk}$ in association with the $L$-violating
$\mu_l$.}\label{f:pdkloop}
\end{figure}
In the mass basis that we have employed, $\lambda''_{112}$ is the only
independent baryon number violating coupling that can be bounded in
conjunction with $\mu_l$ at the tree level. However, the other
$\lambda''$ couplings also drive proton decay at loop levels, and
therefore are also constrained. This can be seen from the
1-loop diagram of Fig.~\ref{f:pdkloop}, where $h^+$ denotes the
physical charged Higgs, given by
	\begin{eqnarray}
h^+ = \sin\beta \; H_d^+ - \cos\beta \; H_u^+ \,,
	\end{eqnarray}
where $\tan\beta\equiv\langle H_u^0 \rangle/\langle H_d^0 \rangle$.
Let us consider the case of a neutrino in the final state. In this
case, the quark on the line $q_k$ must be a down-type quark. The index
$k$ cannot be 3 since there cannot be a $b$-quark in the final
state. For $k=1$ and 2, the final states will be $\pi^+\bar\nu$ and
$K^+\bar\nu$.  Indeed, with $k=1$ and 2, we can cover all the
independent $\lambda''_{ijk}$-couplings because of the antisymmetry of
these couplings in the last two indices. The effective coupling for
the dimension-6 operator is given by
	\begin{equation} 
G'_{\Rsl} \simeq  \left( {f_u^i f_d^j \over
16\pi^2} \right) \left( V_{i1}^* V_{1j} \right)
\frac{\lambda''_{ijk} f_d^k} 
{m^2_{\tilde{d}_R}} 
\left(\frac{\mu^2}{m_{\tilde{H}}^2}\right) \epsilon_l \,,
	\end{equation}
where $\tilde H$ is a linear combination of $\tilde H_u$ and $\tilde
H_d$. 
Although this diagram exists for $\lambda''_{112}$ as well, more
stringent bounds on $\lambda''_{112}\epsilon_l$ originate from the tree
level diagram discussed earlier. For all other combinations, the
bounds are displayed in Table~\ref{t:bounds}.
\begin{table}[b]
\caption[]{\small\sf Summary of upper bounds on the products
$\lambda''_{ijk}\epsilon_l$ derived in this paper. The index $i$
appears as 
row-headings. The indices $j,k$ appear as column headings. The table
entry gives the upper bound, which are independent of the index
$l$, for superparticle masses of order 1\,TeV.\label{t:bounds}} 
\begin{center}
\begin{tabular}{r|ccc}
& 21 & 31 & 32 \\ 
\hline
1 & $10^{-21}$ & $10^{-10}$ & $10^{-11}$\\ 
2 & $10^{-13}$ & $10^{-12}$ & $10^{-13}$\\
3 & $10^{-14}$ & $10^{-13}$ & $10^{-14}$\\
\hline 
\end{tabular}
\end{center}
\end{table}

Before we conclude, a few comments are in order. Although it has been
argued \cite{HaSu} that the $L_iH_u$ terms can be rotated away by
suitable redefinitions of the $L_i$ and $H_d$ fields in the
superpotential, they reappear through quantum corrections after
supersymmetry is broken. Moreover, $\lambda$- and $\lambda'$-couplings
generate bilinear terms via one-loop graphs
\cite{RoMu}. Phenomenological constraints on $\mu_l$ originate from
the fact that the bilinear terms trigger sneutrino-Higgs and
consequently neutrino-neutralino mixing, leading to non-zero neutrino
masses \cite{deC}. Similarly, individual constraints on various
$\lambda''$ exist. They arise, e.g., from $n$-$\bar n$ oscillation
\cite{nnbar} and precision electroweak observables~\cite{bcs}.

To conclude, we have used the constraints on proton stability to
obtain stringent upper limits on all products of the form
$\lambda''_{ijk}\mu_l$. These are complementary to the bounds on the
products of the $L$- and $B$-violating couplings obtained before. The
combinations $\lambda''_{112}\mu_l$ contribute to proton decay at the 
tree level and are therefore very suppressed. The other combinations
drive proton decay through loop diagrams and the resulting constraints
are also displayed. Most importantly, the parameter space we have
examined in this paper has never been explored before.


\begin{thebibliography}{[W]}

\bibitem{rpar} G. Farrar and P. Fayet, {\em Phys. Lett.} {\bf B76}
(1978) 575; S. Weinberg, {\em Phys. Rev.} {\bf D26} (1982) 287;
N. Sakai and T. Yanagida, {\em Nucl. Phys.} {\bf B197} (1982) 533;
C. Aulakh and R. Mohapatra, {\em Phys. Lett.} {\bf B119} (1982) 136;
L. Hall and M. Suzuki, {\em Nucl. Phys.} {\bf B231} (1984) 419; J. Ellis
{\em et al.}, {\em Phys. Lett.} {\bf B150} (1985) 142; G. Ross and
J. Valle, {\em Phys. Lett.} {\bf B151} (1985) 375; S. Dawson,
{\em Nucl. Phys.} {\bf B261} (1985) 297; R. Barbieri and A. Masiero,
{\em Nucl. Phys.} {\bf B267} (1986) 679.

\bibitem{review} For reviews, see G. Bhattacharyya,
\texttt{hep-ph/9709395}, {\em Nucl. Phys. Proc. Suppl.} {\bf 52A}
(1997) 83; H. Dreiner, \texttt{hep-ph/9707435}.

\bibitem{HK} I. Hinchliffe and T. Kaeding, {\em Phys. Rev.} {\bf D47}
(1993) 279.

\bibitem{SmVi98} A.Y. Smirnov and F. Vissani, {\em Phys. Lett.} {\bf
B380} (1996) 317. See also, C. Carlson, P. Roy and M. Sher,
{\em Phys. Lett.} {\bf B357} (1995) 99.

\bibitem{LoPa97} H. N. Long, P.B. Pal, \texttt{hep-ph/9711455}.

\bibitem{pdg} Review of Particle Physics, {\em Phys. Rev.} {\bf D54}
(1996) 1. 

\bibitem{HaSu} Hall and Suzuki, Ref.~\cite{rpar}.

\bibitem{RoMu} S. Roy and B. Mukhopadhyaya, {\em Phys. Rev.} {\bf D55}
(1997) 7020.

\bibitem{deC} F. de Campos, M.~A. Garcia-Jareno, A.~S. Joshipura,
J. Rosiek, J.~W.~F. Valle, {\em Nucl. Phys.} {\bf B451} (1995) 3.

\bibitem{nnbar} F. Zwirner, {\em Phys. Lett.} {\bf B132} (1983) 103;
J.~L. Goity and M. Sher, {\em Phys. Lett.} {\bf B346} (1995) 69,
Erratum {\em ibid} {\bf B385} (1996) 500.

\bibitem{bcs} G. Bhattacharyya, D. Choudhury, K. Sridhar, {\em
Phys. Lett.} {\bf B355} (1995) 193.

\end{thebibliography}
\end{document}